\documentclass[aps,prl,twocolumn,showpacs]{revtex4}%
\usepackage{amsmath}
\usepackage{amsfonts}
\usepackage{amssymb}
\usepackage{graphicx}%
\providecommand{\U}[1]{\protect\rule{.1in}{.1in}}

\begin{document}
\title{Zooming into the coexisting regime of ferromagnetism and superconductivity in
ErRh$_{4}$B$_{4}$ single crystals}
\author{Ruslan Prozorov}
\email{prozorov@ameslab.gov}
\affiliation{Ames Laboratory and Department of Physics \& Astronomy, Iowa State
University, Ames, IA 50011}
\author{Matthew D. Vannette}
\affiliation{Ames Laboratory and Department of Physics \& Astronomy, Iowa State University,
Ames, IA 50011}
\author{Stephanie A. Law}
\affiliation{Ames Laboratory and Department of Physics \& Astronomy, Iowa State University,
Ames, IA 50011}
\author{Sergey L. Bud'ko}
\affiliation{Ames Laboratory and Department of Physics \& Astronomy, Iowa State University,
Ames, IA 50011}
\author{Paul C. Canfield}
\affiliation{Ames Laboratory and Department of Physics \& Astronomy, Iowa State University,
Ames, IA 50011}

\date{29 June 2007}

\begin{abstract}
High resolution measurements of the dynamic magnetic susceptibility are reported for ferromagnetic re-entrant
superconductor, ErRh$_{4}$B$_{4}$. Detailed investigation of the coexisting regime reveals unusual
temperature-asymmetric and magnetically anisotropic behavior. The superconducting phase appears via a series of
discontinuous steps upon warming from the ferromagnetic normal phase, whereas the ferromagnetic phase develops via a
gradual transition. A model based on local field inhomogeneity is proposed to explain the observations.
\end{abstract}



\maketitle

The coexistence of the long-range magnetic order and superconductivity was
first discussed even before the appearance of the microscopic theory of
superconductivity \cite{Ginzburg1957}. Since then this topic remains one of
the most interesting and controversial in the physics of superconductors with
many reviews and books devoted to the subject
\cite{Bulaevskii1985,Sinha1989,Fischer1990,Maple1995,Kulic2006}. Despite
significant effort in new materials design and discovery, there are only few,
confirmed, ferromagnetic superconductors. Local, full-moment ferromagnetic
superconductors: ErRh$_{4}$B$_{4}$ \cite{Fertig1977} ($T_{FM}\approx0.9$ K,
$T_{c}\approx8.7$K), Ho$_{x}$Mo$_{6}$S$_{8}$ \cite{Ishikawa1977}
($T_{FM}\approx0.7$ K, $T_{c}\approx1.8$ K); the weakly ferromagnetic
ErNi$_{2}$B$_{2}$C \cite{Canfield1996} ($T_{FM}\approx2.3$ K, $T_{AFM}%
\approx6$ K, $T_{c}\approx11$ K); and more recent itinerant superconducting
ferromagnets, UGe$_{2}$ ($T_{FM}\approx33$ K, $T_{c}\approx0.95$ K at
$P\approx1.3$ GPa) \cite{Saxena2000} and URhGe ($T_{FM}\approx9.5$ K,
$T_{c}\approx0.27$ K). Whereas all these materials are very interesting on
their own, the coexistence of growing, large, local moment ferromagnetism and
superconductivity is most clearly presented in ErRh$_{4}$B$_{4}$, which is the
subject of the present Letter. In particular, we are interested in the details
of the narrow temperature interval ($\sim0.3$ K) where the two phases coexist
and influence each other.

ErRh$_{4}$B$_{4}$ was extensively studied over past 30 years
\cite{Bulaevskii1985,Sinha1989,Fischer1990,Maple1995,Kulic2006}. The
ferromagnetic phase is primitive tetragonal with the $c-$ axis being the hard
and $a-$ axis being the easy magnetic axes. Detailed measurements of
anisotropic magnetization and upper critical filed, $H_{c2}$ were done by
Crabtree \textit{et al.} \cite{Crabtree1982,Crabtree1986},  who found that
$H_{c2}^{a}$ (along $a-$ axis) peaks at 5.5 K due to large paramagnetic spin
susceptibility in that direction \cite{Crabtree1982,Fischer1990}. In the
contrast, $H_{c2}^{c}$ collapses near the onset of the long-range
ferromagnetic order.

Neutron diffraction studies have established the existence of a modulated
ferromagnetic structure at the lengthscale of $\sim10$ nm
\cite{Moncton1980,Sinha1982}. In single crystals, results suggested that
coexisting phases consists of a mosaic of normal FM domains and SC regions
larger than $\sim200$ nm in size. The SC regions contain modulated FM moment
with a period of $\sim10$ nm. These regions could be regular domains,
spontaneous vortex lattices or laminar structures with $\geq200$ nm
periodicity and modulated SC domains in between \cite{Sinha1982}. Thermal
hysteresis is observed both in the normal Bragg peak intensity and the
small-angle peaks. For the small-angle peaks, the intensity is higher on
cooling than on warming. This is opposite to the behavior of the regular Bragg
peaks from the FM regions \cite{Moncton1980}. Furthermore, the first-order
transition, observed in satellite peaks temperature dependence
\cite{Sinha1982}, is consistent with the spiral state of Blount and Varma
\cite{Blount1979}. However, a continuos transition was reported in other
neutron diffraction \cite{Moncton1977,Moncton1980} and specific heat
experiments \cite{DePuydt1986}. Such a transition can be realized in a
modulated structure or via spontaneous vortex phase.

Theoretically, some striking features of the coexisting phase include an
inhomogeneous, spiral, FM structure \cite{Blount1979,Matsumoto1979} or a fine
domain, "cryptoferromagnetic" phase \cite{Anderson1959,Bulaevskii1985}, a
vortex - lattice modulated spin structure \cite{Tachiki1979}, type-I
superconductivity \cite{Tachiki1979,Gray1983,Bulaevskii1985}, a gapless regime
and possibly, an inhomogeneous Fulde-Ferrell-Larkin-Ovchinnikov (FFLO) state
\cite{Bulaevskii1985}. Another interesting possibility is the development of
superconductivity at the ferromagnetic domain walls
\cite{Buzdin1984,Buzdin2003}.

In this Letter we report precision measurements of the dynamic magnetic
susceptibility of ErRh$_{4}$B$_{4}$ with an emphasis on the narrow temperature
region where ferromagnetism and superconductivity coexist. We find that the
transition is highly asymmetric when FM $\rightarrow$ SC (heating) and SC
$\rightarrow$ FM (cooling) data are compared. The FM$\leftrightarrow$SC
transition proceeds via a series of discrete steps from FM to SC phase upon
warming and proceeds via a smooth crossover from the SC to FM state upon
cooling. With this new information we analyze relevance of some predictions
made over years for the coexisting phase.

Single crystals of ErRh$_{4}$B$_{4}$ were grown at high temperatures from
molten copper flux as described in \cite{Okada1996,Shishido1997}. Resulting
samples were needle shaped with crystallographic c-axis along the needle.
Transport measurements gave residual resistivity ratio of about 8, consistent
with previous reports. The anisotropic $H_{c2}\left(  T\right)  $ curves (see
inset to Fig.\ref{fig3} below) are consistent with earlier reports as well
\cite{Crabtree1982,Crabtree1986}.

The $AC$ magnetic susceptibility, $\chi$, was measured with a tunnel-diode
resonator (TDR) which is sensitive to changes in susceptibility $\Delta
\chi\sim10^{-8}$. Details of the measurement technique are described elsewhere
\cite{Prozorov2000,Prozorov2000a,Prozorov2006}. In brief, properly biased
tunnel diode compensates for losses in the tank circuit, so it is
self-resonating on its resonant frequency, $\omega=1/\sqrt{LC}\sim10$ MHz. A
sample is inserted into the coil on a sapphire rod. The effective inductance
changes and this causes a change in the resonant frequency. This frequency
shift is the measured quantity and it is proportional to the sample dynamic
magnetic susceptibility, $\chi$ \cite{Prozorov2000,Prozorov2000a,Prozorov2006}%
. Knowing geometrical calibration factors of our circuit, we obtain
$\chi\left(  T,H\right)  $. Advantages of this technique are: very small AC
excitation field amplitude ($\sim20$ mOe), which means that it only probes,
but does not disturb the superconducting state; high stability and excellent
temperature resolution ($\sim1$ mK), which allowed detailed study of the
coexisting region, which is only $\sim$ 500 mK wide. Normal-state skin depth
is larger than the sample size, so we probe the entire bulk in the coexisting
region, but when superconducting phase becomes dominant, there is a
possibility that some FM\ patches still exist, but are screened.
\begin{figure}
[ptb]
\begin{center}
\includegraphics[
height=8.0462cm,
width=9.1028cm
]%
{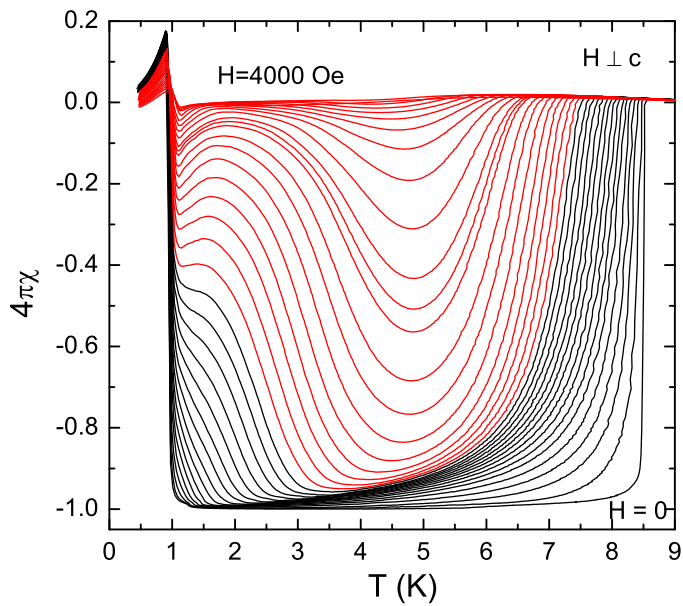}%
\caption{Dynamic magnetic susceptibility in ErRh$_{4}$B$_{4}$ single crystal
measured along the magnetic easy axis (perpendicular to the needle-shaped
sample). Each curve corresponds to a fixed value of the applied dc magnetic
field in the range indicated in the figure. (color online)}%
\label{fig1}%
\end{center}
\end{figure}

Figure \ref{fig1} shows the full temperature scale magnetic susceptibility,
$\chi$, in single crystal ErRh$_{4}$B$_{4}$ for an applied field oriented
along the easy axis and perpendicular to the needle-shaped crystal. The peak
in $\chi\left(  T\right)  $ at the ferromagnetic to superconducting boundary
below $1$ K is a common feature observed in local moment ferromagnets
\cite{Vannette2007}. Clearly, superconductivity is \textit{fully} suppressed
in the ferromagnetic phase. Note that at elevated fields, the response is
nonmonotonic on the SC side close, to the FM boundary, indicative of enhanced
diamagnetism (larger, negative $\chi$), which may be due to suppressed
magnetic pairbreaking or entering into another phase, such as FFLO
\cite{Bulaevskii1985}.%

\begin{figure}
[ptb]
\begin{center}
\includegraphics[
height=7.4026cm,
width=9.1028cm
]%
{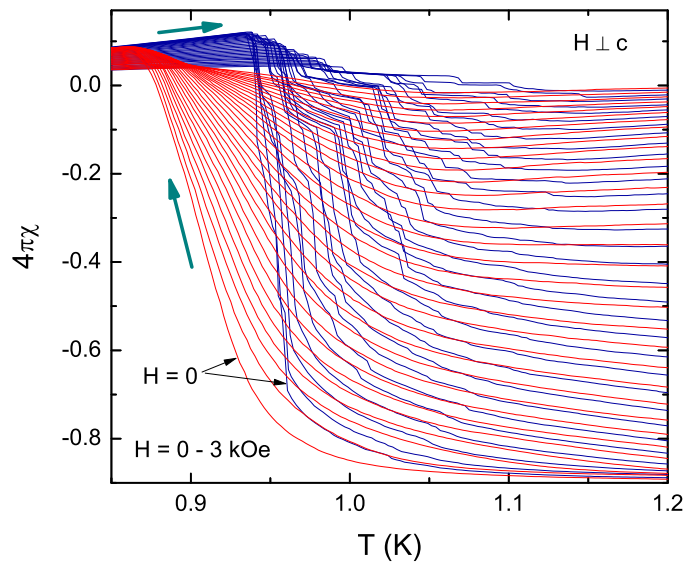}%
\caption{Magnetic susceptibility in a ferromagnet-superconductor transition
region measured at different applied fields. Note the temperature scale and
highly assymetric character of the FM-SC and SC-FM transitions. (color
online). Blue - warming, red - cooling.}%
\label{fig2}%
\end{center}
\end{figure}

Figure \ref{fig2} zooms into the SC$\leftrightarrow$FM transition region.
Measurements were taken after zero-field cooling, applying external field and
warming up (ZFC-W) above $T_{c}$ and then cooling back to the lowest
temperature (FC-C). There is striking asymmetry of the transitions - when the
superconducting phase develops out of the FM state, the response proceeds with
jumps in the susceptibility, which are clearly associated with the appearance
of superconducting regions of finite size. The steps are present up to the
largest field at which superconductivity survives. Decreasing temperature
shows a completely different result: the transition is smooth and gradual and
proceeds to lower temperatures.%

\begin{figure}
[ptb]
\begin{center}
\includegraphics[
height=7.3565cm,
width=9.1028cm
]%
{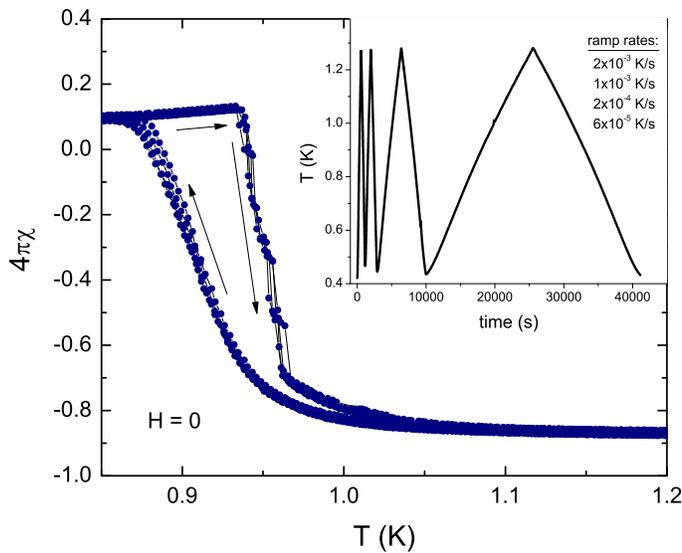}%
\caption{Hysteresis of the transition at zero applied field measured at
different ramp rates. Temperature sweep profiles are shown in the inset.
Clearly, this hysteresis and jumps are not dynamic effects.}%
\label{fig3}%
\end{center}
\end{figure}

To better understand the dynamics of the transition, the top frame of
Fig.~\ref{fig3} shows measurements at $H=0$ for different temperature ramp
rates. Temperature variation is shown in the inset. These data clearly
demonstrate that this hysteresis is insensitive to heating/cooling rates. It
should be noted that all the other data presented in this Letter were taken
with the slowest cooling rate of 60 $\mu$K/s.%

\begin{figure}
[ptb]
\begin{center}
\includegraphics[
height=7.1588cm,
width=9.1028cm
]%
{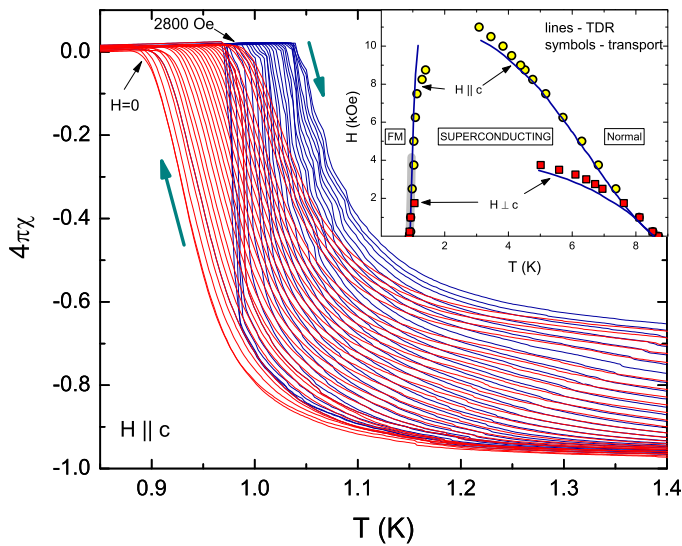}%
\caption{Transition region measured for magnetic field oriented along the
needle and c-axis. Inset: summary phase diagram for two orientations measured
by transport (symbols) and TDR.}%
\label{fig4}%
\end{center}
\end{figure}

Similar hysteresis and steps are also present in the another orientation, when
external magnetic field is applied along the $c-$axis. This is shown in
Fig.~\ref{fig4}. Note that peak in $\chi\left(  T\right)  $ at the
FM\ boundary is not present, which is consistent with the behavior of
anisotropic local-moment ferromagnet \cite{Vannette2007}. The inset to
Fig.~\ref{fig4} shows the phase diagram obtained from resistivity and TDR
measurements for both orientations. There is excellent agreement between the
two techniques and, as noted earlier, this diagram is consistent with previous
reports \cite{Crabtree1982,Crabtree1986}.%

\begin{figure}
[ptb]
\begin{center}
\includegraphics[
height=7.4444cm,
width=9.1028cm
]%
{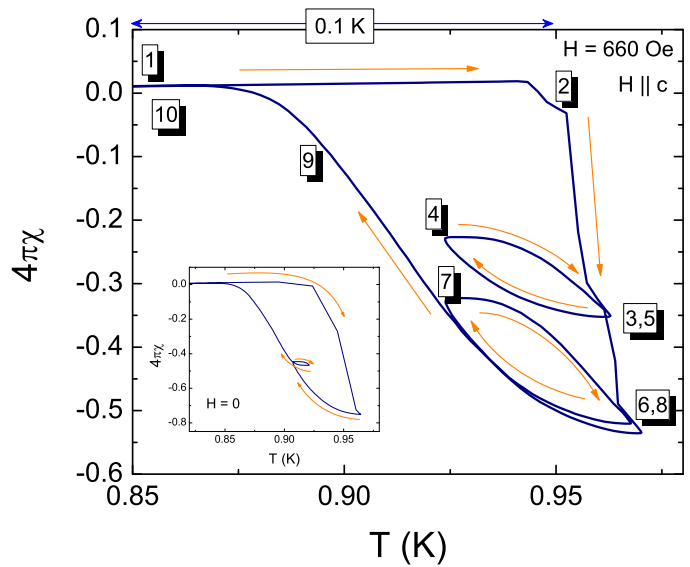}%
\caption{Details of the hysteresis with partial scans as explained in the
text. Inset shows anothe partial loop on a cooling part of the curve.}%
\label{fig5}%
\end{center}
\end{figure}

Finally, Fig.~\ref{fig5} shows so called minor hysteresis loops (not as
function of field, but temperature). The labels show the evolution of the
susceptibility. It starts from low temperature at $(1)$ when sample was warmed
up to first signs of superconductivity that appeared as small jump at $(2)$,
then warmed further and reaching almost full superconductivity at $(3)$, but
then cooled back down to $(4)$ as indicated by arrow and warmed back to $(5)$.
Note that along $(3)\rightarrow(4)$ $\chi$ is significantly different from
$(4)\rightarrow(5)$. Another similar minor cooling-warming loop follows
$(6)\rightarrow(7)\rightarrow(8)$ after which the sample was cooled down to
return to $(10)=(1)$ via $(9)$. Interestingly, there are no steps or jumps
observed on the minor loops even on warming. Also, the slope $d\chi/dt$ is
similar on cooling and warming and is very different from the original steep
slope $(2)\rightarrow(3)$. This is consistent with the present of vortices,
probably pinned by the modulated FM/SC structure. The inset in Fig.~\ref{fig5}
shows a small minor loop on a cooling part. This loop has small slope
comparable to the larger loops described in the main frame.

Let us now turn to the interpretation of these results. Clearly, the
FM$\rightarrow$SC transition proceeds via a series of jumps in diamagnetic
screening due to formation of superconducting regions of macroscopic volume,
roughly $5\%-20\%$ of the sample volume depending on the applied field and
temperature. Indeed, each observed step may be a result of simultaneous
formation of many individual superconducting domains of similar size. These
steps in $\chi$ are present both for $H||c$ and $H||a$ axes, although in the
latter case the steps are smaller and are more pronounced, possibly due to
magnetic and shape anisotropies. The number of steps increases with the
increasing field and the first step (the first sign of superconductivity)
occurs at a higher temperature for larger applied field. Overall,
FM$\rightarrow$SC transition is apparently of the first order and exhibits
behavior consistent with type-I superconductivity as predicted theoretically
\cite{Tachiki1979,Gray1983,Bulaevskii1985}.

From our point of view, the first jump occurs at a temperature where internal
field is equal to a supercooling field of a type-I superconductor. (Note that
apparent superheating of the FM$\rightarrow$SC transition \cite{DePuydt1986}
corresponds to supercooling of a regular type-I superconductor
\cite{Feder1966,DePuydt1986}, because we enter normal phase on cooling). When
first superconducting domains appear, effective magnetic field around them
increases due to the flux expulsion and internal field becomes more
inhomogeneous. This net increase in the internal field in the remaining
FM\ regions stabilizes them to higher temperatures. The system now needs to
get farther away from the initial FM boundary, deeper into the SC state to
produce more superconducting patches. In this scenario, the observed jumps in
$\chi$ correspond to a cascade of supercooling transitions. If the temperature
is lowered before the transition is complete domains stay stable to lower
temperatures due to physics similar to superheating of a type-I
superconductor. It is also quite possible that superconducting domains have
the modulated spin structure seen in neutron scattering \cite{Sinha1982}.
Finally, it seems that ferromagnetic domains are not directly related to the
observed steps, because at higher fields, the number of these domains decrease
and dominant domains (along the applied field) grow in size.

In a striking contrast with FM$\rightarrow$SC transition, the SC$\rightarrow
$FM transition is smooth and proceeds to much lower temperatures. Yet the
transition is hysteretic as evident from the minor loops shown in
Fig.~\ref{fig5}. It is possible that Abrikosov vortices are being
spontaneously created as the temperature is lowered and the systems crosses
over into the normal state when vortex cores overlap. The vortex state is also
compatible with long-range coherence observed in neutron scattering
experiments \cite{Sinha1982}. At the same time ferromagnetic modulation with a
period of $\sim10$ nm may also develop between the vortices
\cite{Anderson1959,Bulaevskii1985,Blount1979,Matsumoto1979}. This would be
also be similar to FFLO state in the presence of vortices \cite{Ichioka2007}.
Moreover, this would explain different intensities of small-angle satellite
peaks, because coherence volume in the vortex state must be much larger
compared to domain-like state on warming. We also note that unusual
enhancement of diamagnetism in the vicinity of the FM boundary from the SC
side, could be due to an FFLO pocket as predicted by Bulaevskii for ErRh$_{4}%
$B$_{4}$ \cite{Bulaevskii1985}. If we plot temperature of a minimum in
$\chi\left(  T\right)  $ as function applied field, and also $H_{c2}$, we
obtain phase diagram remarkably similar to Fig.7 of Ref.\cite{Bulaevskii1985}.

Overall, we conclude that we are witnessing an unusual transition. It is
definitely first order on warming with signs of type-I superconductor
"supercooling". However, it is smooth upon cooling and exhibits smooth minor
loops similar to a type-II superconductor. A second order transition occurs
between normal and FFLO phases as well as between normal and SC for type-II
superconductor. However, transition from SC to FFLO state is first order as
well as from SC to spiral state. It is possible that size of the new phase
nuclei is so small that we cannot resolve it upon cooling.

\acknowledgements{Discussions with Lev Bulaevskii, Alexander Buzdin, Vladimir Kogan and Roman Mints are appreciated.
Work at the Ames Laboratory was supported by the Department of Energy-Basic Energy Sciences under Contract No.
DE-AC02-07CH11358. R. P. acknowledges support from NSF grant number DMR-05-53285 and the Alfred P. Sloan Foundation.}

\bibliography{ErRh4B4}

\end{document}